# METAMATERIAL-INSPIRED WEARABLE PAD FOR ENHANCING EM COUPLING WITH BIOLOGICAL TISSUES

Maria Koutsoupidou[*], Dimitrios C. Tzarouchis[*], *Member, IEEE*, Dionysios Rompolas, Ioannis Sotiriou, George Palikaras, and Panagiotis Kosmas, *Senior Member, IEEE*

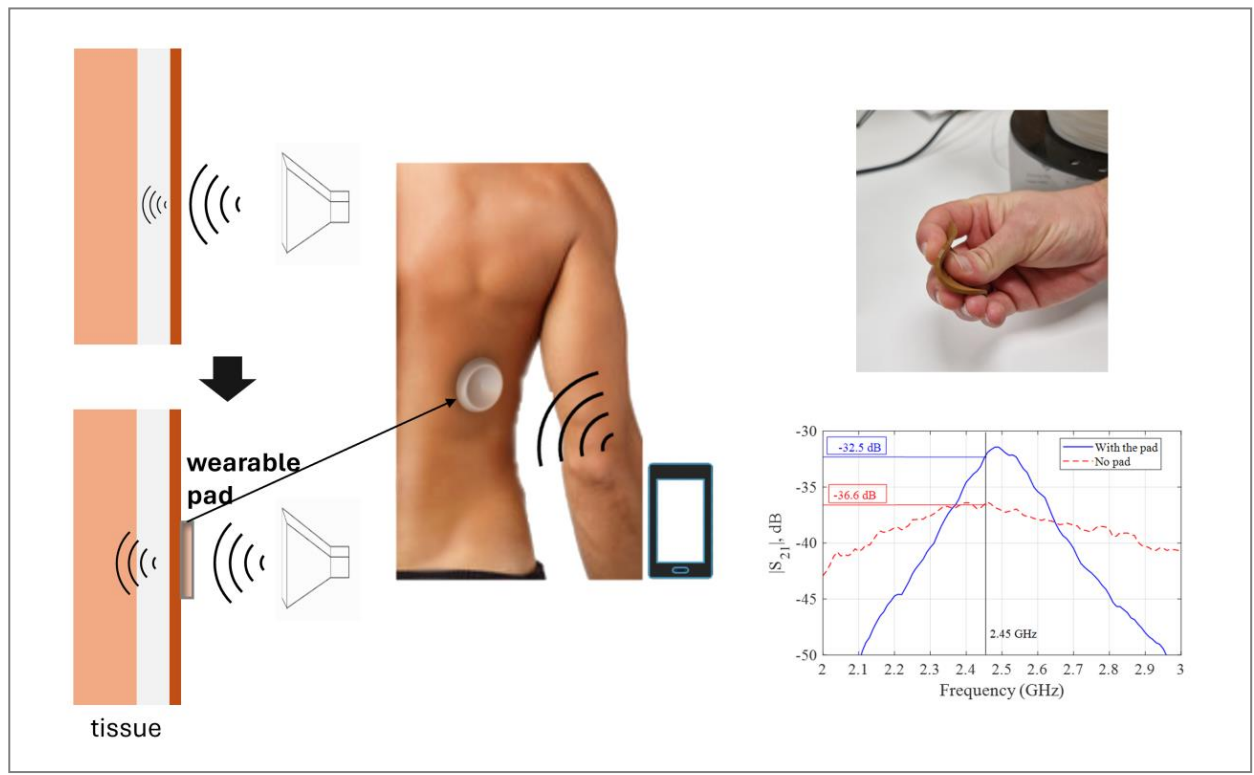

Overview of the functionality of the proposed metamaterial-inspired wearable pad for enhancing the EM coupling with biological tissues at 2.4 GHz.

**Take-Home Messages**

- This paper proposes a totally passive, wearable, thin pad, comprising metallic structures embedded in a dielectric material for effectively coupling EM radiation to a biological tissue.
- The improvement in transmission to a small, implanted antenna was measured to be more than 4 dB in the 2.4-2.5 GHz range.
- The proposed pad can be used as an auxiliary tool in various biomedical applications, from hyperthermia to implant technology for enhancing communication and/or wireless charging.
- A thin, flexible and mechanically robust pad was developed for clinical and every-day biomedical applications.

# METAMATERIAL-INSPIRED WEARABLE PAD FOR ENHANCING EM COUPLING WITH BIOLOGICAL TISSUES

Maria Koutsoupidou*, Dimitrios C. Tzarouchis*, *Member, IEEE*, Dionysios Rompolas, Ioannis Sotiriou, George Palikaras, and Panagiotis Kosmas, *Senior Member, IEEE*

*Abstract* Wearable, implantable, and ingestible antennas are continuously evolving in biomedical applications, as they are crucial components in devices used for monitoring and controlling physiological parameters. This work presents an experimentally validated wearable pad which can improve transmission of electromagnetic waves into the human body. This metamaterial-inspired matching pad, which is based on small metallic loops encased in a thin dielectric layer, is mechanically stable, flexible, and passive. As such, the pad can serve as a coupling medium for microwave medical systems and implantable device communication. Operating in the 2.4-2.5 GHz range, the pad demonstrates significant improvement in signal penetration levels (and, hence, depth) into a biological tissue. The study presents design methodology, simulation studies, in-lab development, and experimental characterization of this pad, which can offer a practical solution for enhanced communication and functionality in various medical diagnostic systems.

*Keywords* —antenna measurements, dielectric materials, electromagnetic coupling, implantable biomedical devices, metamaterials, phantoms

## I. INTRODUCTION[1]

BIOMEDICAL applications covering imaging, sensing, and communication across RF to mm-waves necessitate the transmission and reception of electromagnetic (EM) waves to and from human tissues. Typically, one or more antennas are employed in close proximity to or within the human body. Recent progress in the field of biomedical engineering includes the development of systems of wearable, implantable, and ingestible antennas for monitoring and/or controlling various physiological parameters [1], [2], as well as managing or treating symptoms [3], [4], [5]. The seamless functioning of these devices, especially the long-term operation of implantable ones within the human body, relies heavily on their capacity to establish communication with an external base. This communication is essential for tasks such as data sharing, control, and wireless charging. Similarly, the efficiency of other microwave diagnostic and therapeutic applications, including imaging [6] and hyperthermia [7], depends on the enhanced coupling of the EM radiation with the tissue and the localization toward a specific site or target. The impact of these applications on patients' well-being, quality of life, and healthcare systems is very significant.

The widespread clinical adoption of innovative EM diagnostic and therapeutic tools, particularly implantable technology, requires effective coupling with biological tissues. Two main challenges arise in EM signal propagation between the human body and an external source or detector: a) significant reflection at the air-skin interface, and b) rapid signal attenuation within the tissue due to its high losses, resulting in limited penetration depths relative to the application frequency. This limitation is particularly pronounced in the 2.40-2.50 GHz range (2.4 GHz ISM band), which is the frequency band allocated for various communication protocols like WLAN and Bluetooth [8], [9].

Signal attenuation within tissues presents an inherent challenge, and potential solutions include shifting to lower frequencies, targeting shallow tissue layers, or increasing the overall radiated power. However, these options may not always be feasible for the intended application. Conversely, technical interventions aim to mitigate reflections at the air-skin interface through the use of high permittivity dielectrics [10], [11], [12], [13] and metasurfaces [14], [15] as matching mediums. High permittivity dielectrics are designed to align the air impedance with skin impedance, but they typically involve thick and bulky layers, rendering them possibly suitable for clinical medical devices but not for everyday applications like telemetric control or charging of body implants, which require matching layers that are comfortable and user-friendly. An alternative solution, explored by He et al. [16], [17], utilizes metamaterial effective medium techniques. These methods aim to mitigate reflections through the design of an effective medium comprising subwavelength unit cells featuring metallic rings on a dielectric substrate. However, these approaches entail complex optimization and implementation processes involving arrays of such minuscule unit cells.

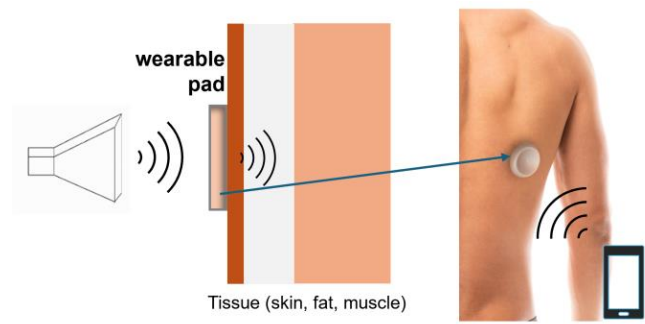

Fig. 1. Concept image of the proposed wearable pad for enhancing implant technology and theranostic systems.

Given these considerations, our paper proposes a matching pad designed to conform to the human body above the targeted area, significantly enhancing the external field's impact on the body. The pad's metamaterial-inspired design incorporates small parallel metallic loops within a thin dielectric of custom permittivity. Upon excitation by an external field, these metal loops induce resonances at frequencies which depend on their dimensions. Moreover, the dielectric cover of these loops acts as a matching medium at the air-skin interface and contributes to the mechanical stability of the setup. This design enables a pad which is mechanically stable, flexible, biocompatible, and entirely passive.

This pad is intended to serve as a coupling medium between incident EM radiation and human tissue for medical diagnostics or therapy systems requiring microwaves, as well as for communication with implantable devices (Fig. 1). The choice of the 2.4 to 2.5 GHz range for the operating bandwidth is specifically tailored to suit the latter application. The upcoming sections will present the design process and simulation outcomes across different setups (Section II), the in-lab development methodology for the pad (Section II), and its experimental characterization (Section III).

## II. METHODS AND PROCEDURES

### A. Wearable pad design and specifications

Our approach in designing the wearable matching pad prioritized simplicity, mechanical feasibility, and robustness. Instead of employing an array of metallic structures [14], [16], [17], we opted for a minimalist configuration consisting of only one set of rings embedded vertically within a medium with low permittivity and minimal losses. By doing so, we reduce the complexity of the problem, minimizing unknowns and eliminating the need for homogenization techniques.

The proposed design features two parallel metal rings encased within a thin dielectric circular pad (Fig. 2). The underlying principle is that induced currents in the metallic rings generate an effective magnetic dipole moment, facilitating the re-radiation of energy into the intricate and lossy biological tissue. The rings were presumed to possess specific thicknesses as they are crafted from commonly used copper sheets or copper tape. Our sole imposed constraint on these rings was to ensure they weren't positioned too close to the surface of the pad and risk to touch the human skin. Their placements were meticulously optimized to yield optimal performance and effectiveness in the intended application.

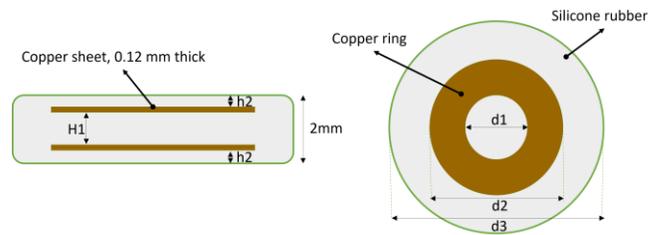

Fig. 2. Side (left) and top (right) view of the design of the matching pad made of two parallel copper rings embedded in a thin silicon rubber disk.

Both the metallic loops and the dielectric pad were inverse designed: the resulted permittivity of the pad and the radii of the circular metallic rings were subject to optimization. The design goal was to achieve maximization of the power transmission from an external source to the tissue. The basic idea emerged from the contemporary trends of inverse design and metamaterial synthesis [18].

The setup was designed to operate at 2.45 GHz and to exhibit a maximum thickness of 2 mm, to ensure that it is comfortable for users and thus useful in practical applications. The metal rings were designed to be made of copper sheet of 0.12 mm thickness inserted into the dielectric pad with dielectric permittivity, $\varepsilon_r = 4$, and loss tangent, $\tan\delta = 0.02$. The rings were positioned flexibly within the pad, ensuring they stayed at least 0.3 mm away from the outer edge.

With these design specifications and constraints, the pad was optimized to maximize the EM field inside the tissue beneath the pad in the 2.4-2.5 GHz range. The tissue was modelled as a three-layer stack of: a) 5 mm thick skin with dielectric permittivity $\varepsilon_{r\_skin} = 39.2$ and conductivity $\sigma_{skin} = 1.8$ S/m, b) 25 mm thick fat with $\varepsilon_{r\_fat} = 5.0$ and $\sigma_{fat} = 0.25$ S/m and c) 30 mm muscle with $\varepsilon_{r\ muscle} = 52.7$ and $\sigma_{muscle} = 1.95$ S/m. The EM field incident to the pad was linearly polarized, generated by a dipole with a resonance at 2.45 GHz. Since the pad's topology is symmetric, the orientation of the field does not affect its performance. The setup was modelled with CST Microwave Studio Suite® and optimized using the CST GA optimization tool. The pad's dimensions were selected as, $H_1 = 1.16$ mm, $h_2 = 0.3$ mm, $d_1 = 15.5$ mm, $d_2 = 25$ mm, and $d_3 = 55$ mm.

## B. Simulation results

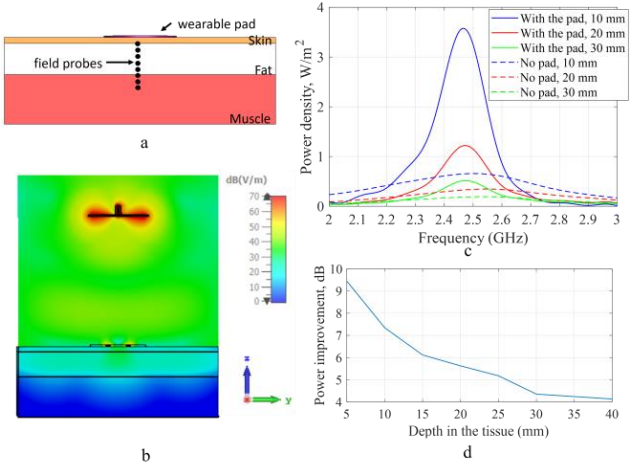

Fig. 4. (a) Field distribution of the electric field excited by a dipole antenna located at 130 mm above a three-layer tissue comprising (from top to bottom) skin, fat, and muscle, with the pad lying on the tissue surface. (b) Power density calculated at depths of 10 mm, 20 mm, and 30 mm below the tissue surface with (solid line) and without (dashed line) the pad in the 2-3 GHz range. (c) Improvement in dB of the power density with the pad for different depths in the tissue at 2.45 GHz.

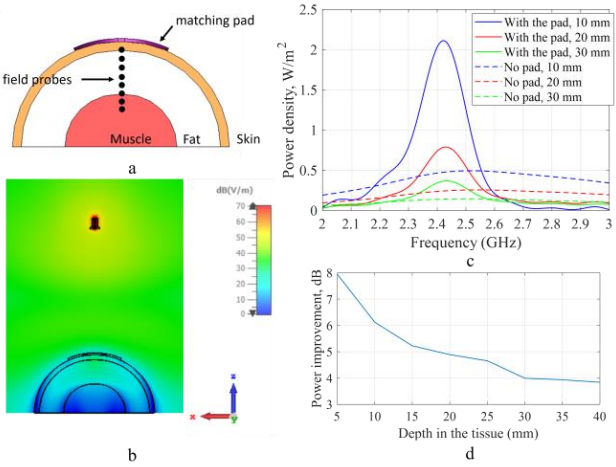

Fig. 3. (a) Simulation setup where the matching pad conforms to tissue of cylindrical shape of 120 mm diameter that comprises skin, fat, muscle. (b) Field distribution of the electric field excited by a dipole antenna 130 mm above the cylindrical tissue and the conforming matching pad. (c) Power density calculated at depths of 10 mm, 20 mm, and 30 mm below the tissue surface with (solid line) and without (dashed line) the pad in the 2-3 GHz range. (d) Improvement in dB of the power density with the pad for different depths in the tissue at 2.45 GHz.

The electric and magnetic fields were monitored with probes at various tissue depths to calculate power density with and without the pad (Fig. 4a). Fig. 4b illustrates the electric field distribution inside the tissue at 2.45 GHz (central frequency) when the pad is employed. The power density near the pad (5 mm deep in the tissue) shows a 9 dB improvement, extending to 4 dB at a depth of 40 mm within the tissue (Fig. 4c, d). Importantly, the polarization of the field inside the tissue remains unaffected by the presence of the pad.

Additionally, the proposed pad should maintain its performance when conforming to curved body parts, such as hands or legs. To access this, the pad was tested on a cylindrical setup of 120 mm diameter comprising the previous setup layers (Fig. 3a). The simulation results depicting the electric field distribution are presented in Fig. 3b. Bending the pad induces a slight frequency shift in its resonance from 2.46 GHz to 2.42 GHz (Fig. 3c). Nevertheless, for both configurations, there is a noticeable enhancement across the entire 2.4 GHz ISM range. When bent, the improvement ranges from 8 dB at 5 mm within the tissue to 4 dB at 40 mm (Fig. 3d). In both cases, strong attenuation is noticeable at a depth of 30 mm (fat-muscle interface) due to the high conductivity of the muscle, resulting in strong signal attenuation. However, the degradation occurs more gradually as the signal continues to propagate deeper into the muscle layer.

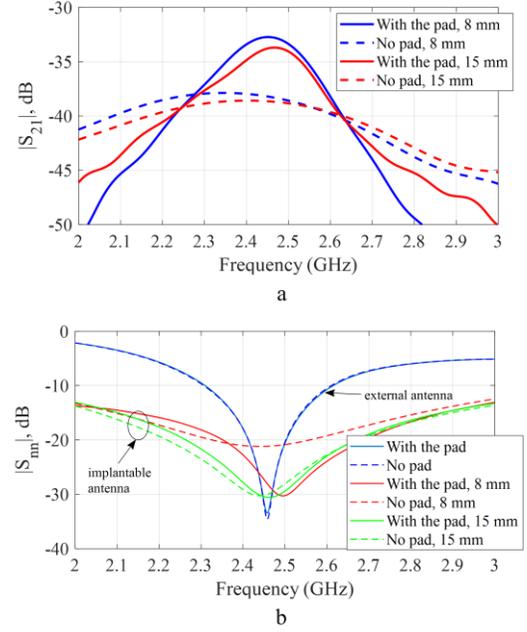

Fig. 5. (a) Transmission coefficient and (b) reflection coefficient of the external antenna and the implantable antenna at depths 8 mm and 15 mm from the tissue surface with (solid line) and without (dashed line) the pad on the tissue surface.

These results clearly demonstrate the enhanced power penetration into the tissue when utilizing the proposed pad. In addition to this, and to consider various microwave applications which involve implantable antennas within the tissue, (e.g., telemetry and wireless charging), we conducted tests on the transmission between an external antenna and a dipole implanted in the fat layer. To achieve this measurement, the previous flat-tissue model was updated to incorporate a small dipole antenna operating at 2.45 GHz, positioned at depths of 8 mm and 15 mm within the fat layer. The magnitude of the transmission coefficients, $|S_{21}|$, for both implant positions, with and without the pad, are depicted in Fig. 5a, while the corresponding reflection coefficients ($|S_{11}|$, $|S_{22}|$) of the external and implanted dipole are presented in Fig. 5b. The transmission enhancement at 2.45 GHz is 5.4 dB and 4.8 dB for positions at 8 mm and 15 mm, respectively. As anticipated and is evident from the reflection coefficient results, the pad influences the antenna's performance when the implant is closer to the tissue's surface. Conversely, the external antenna is positioned far enough to avoid impacting its input resistance.

*C. Wearable pad development*

The pad is composed of copper rings embedded in a dielectric material, for which we opted for a blend of silicon rubber with calcium copper titanite (CCTO). Silicon rubber was chosen for its mechanical strength and flexibility [19], while CCTO powder was introduced to further control the dielectric permittivity and reduce the loss tangent of the mixture. The copper rings, crafted from copper sheet cut with a laser cutter, were intentionally designed to be fully immersed in the pad, ensuring no contact with the user's skin.

For the mixture, the silicon rubber base and hardener (Zhermack GmbH, Germany) and CCTO powder (Lori Industry Co., Ltd., China) were blended in a ratio of 100:3:200 by mass. In its semi-solid state, the mixture underwent degassing in a vacuum chamber to eliminate air bubbles. Subsequently, its dielectric properties were measured within the 2-3 GHz range using the open-ended coaxial probe DAK 12 by SPEAG® (Fig. 6a). The semi-solid mixture was then poured into a 3D printed mold and left to cure. The resulting pad exhibits excellent mechanical stability and flexibility, enabling it to conform to various body parts. Its overall dimensions are 2.13 mm in thickness and 54.6 mm in diameter (Fig. 6b). Moreover, it is lightweight, weighing only 8 g.

*D. Experimental setup*

To test the developed pad and validate the simulation results, we constructed the final simulation setup, incorporating a dipole antenna positioned within the fat layer of a flat phantom. We assembled a box with acrylic walls and 3D printed components (Fig. 7a), including sockets of varying heights designed to accommodate the implanted dipole. Specifically, two sockets were made: one for placing the antenna at a depth of 8 mm below the surface and another at 15 mm (Fig. 7b).

The muscle, fat and skin gel phantoms were made with the following process. For the muscle and skin phantoms, DI water, gelatin and glycerol were used. For the skin, gelatin and water were mixed in a mass ratio 1/20 at 75°C until the mixture was homogenous, and subsequently, glycerol was gradually added in the mixture until the desired dielectric properties were reached by continuously measuring its dielectric properties. Finally, the mixture was molded and placed in the fridge to solidify for a few hours. Similarly, DI water, gelatin, safflower oil and wheat flour were used to fabricate the fat phantom. Gelatin and water were again initially mixed, and then flour was added in the mixture at a gelatin/flour mass ratio equal to 1/5. Finally, safflower oil was gradually added until the desired dielectric properties were reached, again, by continuously measuring mixture properties. The mixture was poured in the mold and placed in the fridge.

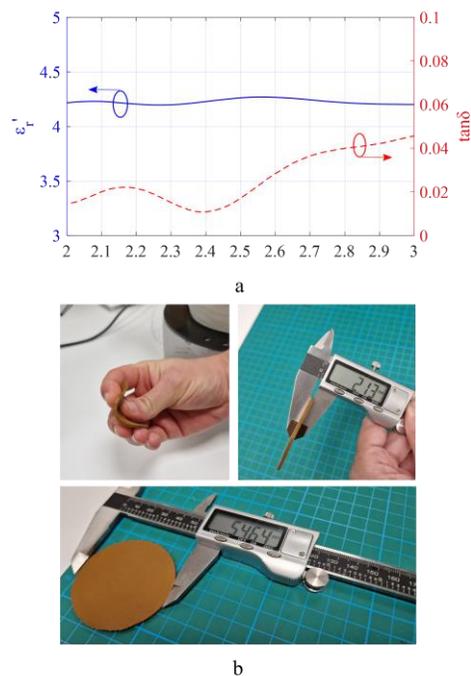

Fig. 6. (a) Transmission coefficient and (b) reflection coefficient of the external antenna and the implantable antenna at depths 8 mm and 15 mm from the tissue surface with (solid line) and without the pad (dashed line) on the tissue surface.

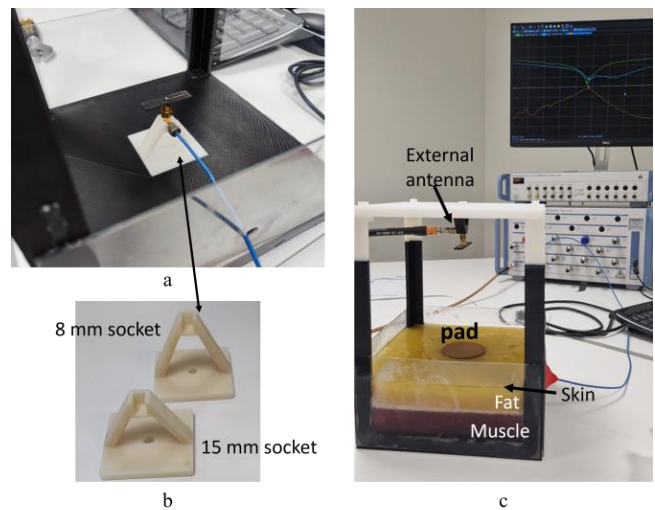

Fig. 7. (a) Transmission coefficient and (b) reflection coefficient of the external antenna and the implantable antenna at depths 8 mm and 15 mm from the tissue surface with (solid line) and without the pad (dashed line) on the tissue surface.

TABLE I
DIELECTRIC PROPERTIES OF THE TISSUE MODELS AND PHANTOMS

| 2.45 GHz | Simulation model | | Experimental setup 1 (antenna at 8 mm) | | Experimental setup 1 (antenna at 15 mm) | |
|---|---|---|---|---|---|---|
| | $\varepsilon_r$ | $\sigma$ (S/m) | $\varepsilon_r$ | $\sigma$ (S/m) | $\varepsilon_r$ | $\sigma$ (S/m) |
| skin | 39.2 | 1.8 | 39.7 | 3.9 | 38.8 | 3.1 |
| fat | 5 | 0.25 | 5.7 | 0.16 | 5.2 | 0.14 |
| muscle | 52.7 | 1.95 | 52.7 | 1.85 | 53.4 | 2.35 |

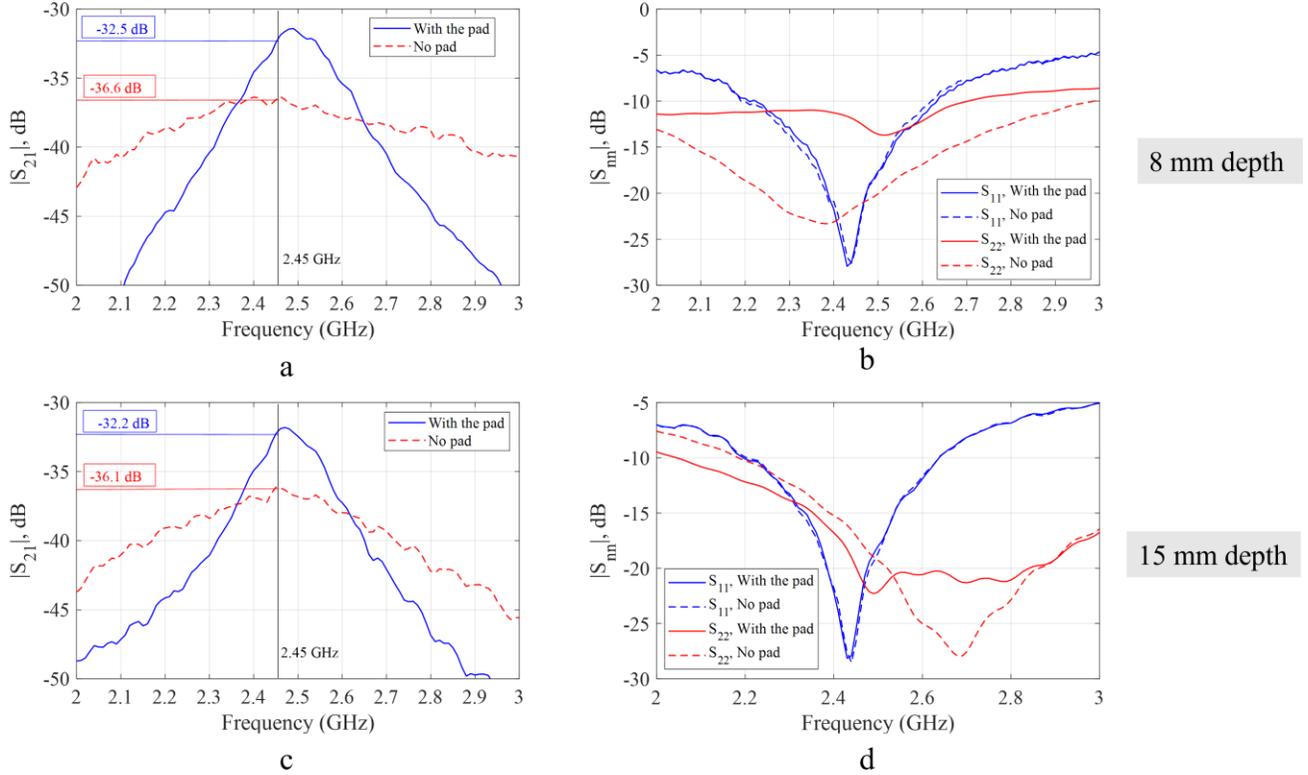

Fig. 8. (a) Magnitude of transmission coefficient, |S21|, and (b) Magnitude the reflection coefficients of the external and the immersed dipole, |S11|, |S22|, with (solid line) and without (dashed line) the pad when the immersed dipole lied 8 mm below the phantom surface. (c) Magnitude of transmission coefficient, |S21|, and (d) Magnitude the reflection coefficients of the external and the immersed dipole, |S11|, |S22|, with (solid line) and without (dashed line) the pad when the immersed dipole lied 15 mm below the phantom surface.

The dielectric properties of the developed phantoms measured at 2.45 GHz with the DAK12 probe (SPEAG®) and the results vs the model values are shown in Table I. Initially, the socket for 8 mm was used and each gel mixture was poured following solidification of the previous one. The final experimental setup is shown in Fig. 7c: the metallic loop-based pad lies on the three-layer phantom and the external antenna is attached to the top cover of the box, to ensure that it is stable and located at specific, controlled distances during the experiment. The antennas' S-parameters with and without the pad were measured using an R&S®ZNBT VNA. To replicate the experiment with the antenna positioned inside the 15 mm socket within the tissue, new phantoms were fabricated after destroying the previous ones, and the experiment was conducted again.

### III. EXPERIMENTAL RESULTS

The experimental results for the S-parameters of the dipole antennas in the 2-3 GHz frequency range for both depths, 8 mm and 15 mm, are depicted in Fig. 8. When the immersed dipole lies at 8 mm from the phantom surface, the magnitude of the transmission coefficient, |S21|, is enhanced by 4.1 dB at 2.45 GHz (Fig. 8a), although this dipole had

TABLE II
TRANSMISSION IMPROVEMENT WHEN THE PAD IS USED AT SELECTED FREQUENCIES

| Depth of implanted antenna | 2.4 GHz | 2.45 GHz | 2.5 GHz |
|---|---|---|---|
| 8 mm | 1.8 dB | 4.1 dB | 5.4 dB |
| 15 mm | 2.0 dB | 3.9 dB | 4.4 dB |

been significantly detuned (Fig. 8b). Additionally, the transmission enhancement because of the pad is higher than 2 dB from 2.4 GHz to 2.6 GHz, peaking at 2.5 GHz. We get similar results for the 15 mm depth, where the pad improved the |S21| at 2.45 GHz by 3.9 dB (Fig. 8c). The immersed dipole is more stable at this depth because of its increased distance from the pad's inductive loops. Generally, the transmission enhancement matches well the simulation results of Section II.B. A summary of the measured improvement in $S_{21}$ scattering at specific frequencies in the 2.4-2.5 GHz frequency range is shown in Table II.

### IV. DISCUSSION AND CONCLUSIONS

Our findings indicate that employing the proposed metal-loop-based pad enhances the penetration depth of incident EM radiation through a combination of dielectric matching mechanisms and the excitation of a resonance near the tissue. We note that the observed improvement of 4-5 dB in transmission to an implant holds exceptional significance, as this difference can often make a signal connection feasible and reduce the power consumption of active implantable devices. In implant technology, overcoming connection gaps is a substantial challenge, and this enhancement addresses that issue effectively. Furthermore, the chosen materials and design at 2.4 GHz ensures comfortable, safe, and user-friendly application to users in everyday life. The pad's small size, light weight, and ability to be attached to any body part using an available adhesion mechanism contribute to its practicality. Additionally, a simulation-based sensitivity analysis was conducted to assess the system's robustness. The $\pm 3\sigma$ analysis revealed that, with a 10% standard deviation ($\sigma = 10\%$) in the electromagnetic properties (permittivity and loss tangent) of both the tissues and the pad, there is a $\pm 1.5$ dB variation in the calculated transmission. Thus, the system demonstrates robustness against perturbations and uncertainties.

Based on these findings, we believe that the proposed pad can tackle critical challenges in the realm of biomedical applications involving the transmission and reception of EM waves within the RF to mm-wave spectrum. Aware of the limitations posed by signal attenuation and reflections at the air-skin interface, we have developed an innovative solution in the form of a metamaterial-inspired matching pad. The next steps in this research involve scaling the proposed design to higher frequency ranges, such as 3.6 GHz (mid-band 5G), and incorporating multispectral characteristics. This expansion aims to enable simultaneous operation across various frequency ranges and communication protocols. Overall, we believe that this work can contribute to the advancement of microwave medical diagnostics and therapy systems, offering promising potential for both clinical and everyday use by patients.